\documentclass[twocolumn,aps,prc,floatfix]
{revtex4}

\usepackage[dvips]{graphicx}         %
\usepackage{bm}                      % bold math
\usepackage{mathptmx}                % math+times fonts
\usepackage{dcolumn}                 % Align table columns on decimal point
\usepackage{multirow}                %
\def\bea{\begin{eqnarray}} \def\eea{\end{eqnarray}}
\def\beq{\begin{equation}} \def\eeq{\end{equation}}
\def\bal#1\eal{\begin{align}#1\end{align}}
\def\bse#1\ese{\begin{subequations}#1\end{subequations}}
\begin{document}

\title{Imprints of high-momentum nucleons in nuclei on hard photons from heavy-ion collisions around the Fermi energy}

\author{Wen-Mei Guo$^1$\footnote{guowenmei@ahnu.edu.cn}}
\author{Bao-An Li$^2$\footnote{Bao-An.Li@TAMUC.edu}}
\author{Gao-Chan Yong$^3$\footnote{yonggaochan@impcas.ac.cn}}
\affiliation{
$^1${Department of Physics, Anhui Normal University, Wuhu 241000, China}\\
$^2${Department of Physics and Astronomy, Texas A\&M University-commerce, Commerce, Texas 75429-3011, USA}\\
$^3${Institute of Modern Physics, Chinese Academy of Sciences, Lanzhou 730000, China}}

\date{\today}

\begin{abstract}
The short-range correlation (SRC) induced by the tensor force in the isosinglet neutron-proton interaction channel leads to a high-momentum tail (HMT) in the single-nucleon momentum distributions n(k) in nuclei. Owing to the remaining uncertainties about the tensor force, the shape of the nucleon HMT may be significantly different from the dilute interacting Fermi gas model prediction $n(k) \sim1/k^4$ similar to the HMT in cold atoms near the unitary limit.
Within an isospin- and momentum-dependent Boltzmann-Uehling-Uhlenbeck transport model incorporating approximately the nucleon HMT, we investigate hard photon emissions in $^{14}$N+$^{12}$C and $^{48}$Ca+$^{124}$Sn reactions at beam energies around the Fermi energy. Imprints of different shapes of the HMT on the energy spectrum, angular distribution and transverse momentum spectrum of hard photons are studied.  While the angular distribution does not carry any information about the shape of the nucleon HMT, the energy spectra and especially the mid-rapidity transverse momentum spectra of hard photons are found to bare strong imprints of the shapes of nucleon HMTs in the two colliding nuclei.
\end{abstract}

%\pacs{21.30.Fe, 21.65.Cd, 26.60.-c}

%\showkeys{Nuclear Interaction, Induced Interaction, Nuclear Matter}

\maketitle

%===============================================================================
\section{Introduction}
The nucleon-nucleon short-range repulsive core (correlations) and
tensor force are known to lead to a high (low) momentum tail (depletion) in the
single-nucleon momentum ($k$) distribution $n(k)$ above (below) the nucleon Fermi momentum $k_F$ \,\cite{Mig57,bethe,pan92,Pan99}.
The study on the shape of the nucleon high momentum tail (HMT) at $k>k_F$ has a long history, see e.g., Refs. \cite{Mig57,Gal58,Lut60,Czy61,Bel61,Sar80,Sar82}.
The HMT can be derived by expanding quantities of interest in a dilute interacting gas of hard spheres in terms of the $k_F\cdot a$ where a is the scattering length or hard-core radius. The asymptotic behavior of $n(k)$ at high momenta is generally related to the Fourier transform $\widetilde{V}_{\text{NN}}(k)$ of the nucleon-nucleon (NN) interaction via ~\cite{Amado76a,Amado76b,Amado77,Ant93}
$
n(k)\rightarrow[k\rightarrow
\infty]{}[\widetilde{V}_{\text{NN}}(k)/k^2]^2.
$
Thus, by investigating the HMT of $n(k)$ one may get invaluable information about the still poorly known nuclear forces at short distances especially the tensor force existing mostly in the isosinglet neutron-proton interaction channel and the resulting nucleon-nucleon short range correlation (SRC), see, e.g., Ref. \cite{Ant93} for a historical review and Ref. \cite{Hen16x} for a more recent review of both theoretical and experimental investigations of the HMT and SRC. Implications of the HMT and SRC on properties of nuclear matter, such as the density dependence of nuclear symmetry energy \cite{Ohen15}, can be found in the recent review in Ref. \cite{PPNP},  while their impacts on properties of neutron stars, such as the tidal deformation, cooling and mass-radius correlation, can be found in Refs. \cite{Cai15,Souza1,Souza20,Lu21}. Despite of the impressive progresses in the field, many interesting questions regarding the nature, size, range and shape of the HMT and SRC need to be more thoroughly studied. For example,  within the neutron-proton dominance model using a contact force, the nucleon HMT naturally reduces asymptotically to the $n(k) \sim1/k^4$ as for the ultra cold atoms near the unitary limit \cite{Hen15}, and it is consistent with predictions of dilute Fermi gas models \cite{Sar80,Sar82,Rios13}.  On the other hand, there are indications from both theories and experiments that
the HMT may not scale as $n(k) \sim1/k^4$. For example, earlier analyses of scaling functions, e.g., y-scaling and superscaling in the quasi-elastic region of inclusive electron-nucleus scatterings \cite{Ant07}, found that the $n(k)$ scales as $n(k) \sim1/k^{4+m}$ with $m\approx 4-4.5$ for momenta $k$ up to $(1.59\sim1.97)k_F$ with $k_F=250$ MeV/c. Different powers in the HMT mean different short range behavior of nuclear forces. For example, the nuclear force from an inverse Fourier transform for $m=4$ and $m=5$ behaves as ${V}_{\text{NN}}(r)\sim 1/r$ and ${V}_{\text{NN}}(r)\sim (1/r)^{1/2}$, respectively.
Therefore, determining the shape of the HMT has important ramifications for our understanding about the nature of strong interactions at short distances.

It is well known that observables of heavy-ion collisions from low to relativistic energies are sensitive to the initial phase space distributions of nucleons or quarks and gluons in the colliding nuclei. For examples, several observables in ultra-relativistic heavy-ion collisions have been used or proposed to probe the deformations and neutron skins of heavy nuclei, see, e.g., Refs. \cite{Li00,HLi20,Jia21,Gia20}. It is also well known that among all observables of heavy-ion reactions, photons are among the most clean and undistorted messengers of the earlier stage of nuclear reactions because they only interact with nuclear matter electromagnetically unlike hadrons which suffer from strong final state interactions on their way to the detector.  While the electromagnetic field created during heavy-ion collisions was recently found to affect appreciably the collectivity of photons, its effects on the photon yields and spectra are relatively small \cite{Ma18}. Moreover, hard photons (with energies above 30 MeV distinctly different from photons emitted from giant dipole
resonances \cite{Giul06}) from heavy-ion collisions are known to be sensitive to the HMT of single nucleon momentum distributions in the colliding nuclei \cite{Bertsch88,Cass90,Bona06,Bona94}.
This is mainly because theoretical studies have consistently indicated hard photons are emitted mainly from incoherent proton-neutron bremsstrahlung $p+n\rightarrow p+n+\gamma$ processes, during the early stage of heavy-ion collisions. Indeed, various theoretical studies incorporating this hard photon production mechanism indicate that hard photons are good probes of the nucleon momentum distributions in the collision zone during the earlier stage of the reaction \cite{Ko85,Cassing86,Nak86,Bauer86,Bauer86a,Rem86,Ko87,Biro87,Niita,Gan94,Yong08,Liu08,Shicz20,WangSS20}. In heavy-ion collisions at beam energies around the Fermi energy, hard photons are mostly from the first chance neutron-proton bremsstrahlungs. They are thus useful for probing the initial HMT in the two colliding nuclei. While it was speculated in earlier studies that the HMT would be important for hard photon productions, it was not considered until recently \cite{Yong17,Yong172,Yong18} thanks to the quantitative information about the size and isospin dependence of the SRC from the new electron-nucleus scattering experiments \cite{Hen16x}. However, only the $n(k) \sim1/k^4$ HMT was used in the studies so far.  Compared to ongoing and planned experiments using electron/nucleon-nucleus scatterings to probe the nature of SRC and the resulting HMT, heavy-ion collisions can use constructively HMT effects in the two colliding nuclei, may thus lead to enhanced SRC/HMT effects.

There are some interests to investigate effects of the shape of the nucleon HMT on hard photon production in heavy-ion reactions at Fermi energies \cite{Alan,Giu}. Besides the theoretical importance of knowing the shape of the nucleon HMT mentioned earlier, one of the goals of this work is to provide theoretical inputs and motivations for future heavy-ion reaction experiments. Within an isospin- and momentum-dependent Boltzmann-Uehling-Uhlenbeck (IBUU04) transport model \cite{LiBA04a,LiBA04b,LiChen05,Liba05} incorporating approximately the nucleon HMT induced by the SRC in colliding nuclei \cite{Yong08,Yong17,Yong172,Yong18}, we investigate hard photon emissions in $^{14}$N+$^{12}$C and $^{48}$Ca+$^{124}$Sn reactions around the Fermi energy. Effects of different shapes of the HMT on the energy spectrum, angular distribution and transverse momentum spectrum of hard photons are studied in $^{48}$Ca+$^{124}$Sn at beam energies of $E_b=30$, 45 and 60 MeV/nucleon. The transverse momentum spectra of hard photons are found to be most sensitive to the shape of HMT.

The rest of the paper is organized as follows. In the next section, we shall summarize the most relevant components of the IBUU04 transport model for simulating heavy-ion reactions at intermediate energies, the modeling of initial nucleon momentum distributions with HMTs and the elementary cross section for the $p+n\rightarrow p+n+\gamma$ process. We then study effects of the HMT on the photon production dynamics, angular and energy spectrum as well as the impact parameter and beam energy dependence of hard photon production with $n(k) \sim1/k^4$. We also try to get a felling about the reliability of the model by comparing our calculations with the old and minimum bias photon spectra from $^{14}$N+$^{12}$C reactions at beam energies of $E_b=20$, 30 and 40 MeV/nucleon. Afterwards, we investigate effects of the shape of HMT on the energy spectrum, angular and transverse momentum distributions of hard photons in $^{48}$Ca+$^{124}$Sn reactions at beam energies of $E_b=30$, 45 and 60 MeV/nucleon using $n(k) \sim1/k^4, 1/k^6$ and $1/k^9$, respectively. Finally, we summarize our main findings.

\section{Theoretical framework}
For completeness and ease of our discussions, we summarize here the most relevant components of our approach.
For simulating the reaction dynamics, we use the updated IBUU04 transport model incorporating approximately the HMT in the colliding nuclei.
Details of the IBUU04 transport model can be found in Refs. \cite{LiBA04a,LiBA04b,LiChen05,Liba05} while the method used to incorporate the HMT can be found in Refs. \cite{Yong08,Yong17,Yong172,Yong18}.
\begin{figure}[t!]%..............................................................
\centering
\setlength{\abovecaptionskip}{-6.cm}
\includegraphics[width=0.9\textwidth]{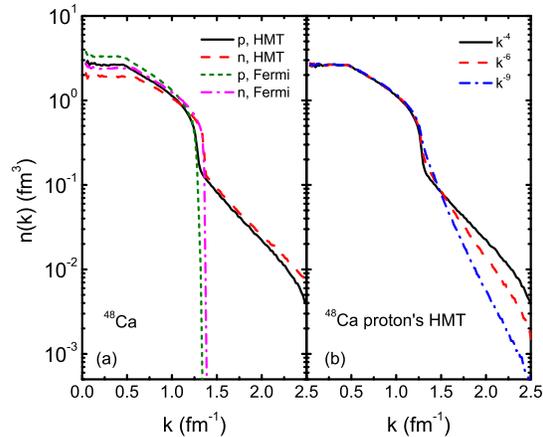}
\caption{Left: Nucleon momentum distributions $n(k)$ in $^{48}_{20}$Ca with a HMT in the shape of $n(k)\sim 1/k^4$. The distributions without the HMT are also shown for a comparison. Right: Momentum distributions for protons with HMT in the form of $n(k)\sim 1/k^4$, $1/k^6$ and $1/k^9$, respectively.} \label{hmts}
\end{figure}

The initial density distributions of
nucleons in the projectile and target are obtained from the Skyrme-Hartree-Fork calculations
with the Skyrme M$^*$ parameter set \cite{Frie86}.
The nucleon momentum distribution is modeled according to~\cite{Ohen15,Yong16,Cai16}
\beq\label{mom}
n(k)=\left\{
\begin{array}{ll}
C_1, &k\leq k_F,\\ C_2/k^{4+m}, &k_F<k<\lambda k_F,
\end{array}
\right.
\eeq
where $k_F$ is the Fermi momentum obtained from the local density using the Thomas-Fermi approximation and
$\lambda\approx2$ is the high-momentum cutoff.
The parameters $C_1$ and $C_2$ are determined by the normalizaion condition of $n(k)$ as well as a specified
fraction of neutrons and protons in their respective HMTs.
Here, $20\%$ of total nucleons are distributed in the HMT according to the experimental findings \cite {Sube08,OHen14}.
We also adopt the n-p dominance model \cite{OHen14} where equal numbers of neutrons and protons are required to be in the HMT.
In this model, the HMT and SRC are caused completely by the tensor force between the isosinglet neutron-proton pairs.
We use $m=0,\ 2$ and $5$ to model different shapes of HMT. Shown in Fig.~1 (a) are the momentum distributions of neutrons and protons  in $^{48}_{20}$Ca
with a HMT in the form of $1/k^4$. For a comparison, the nucleon momentum distributions using the Thomas-Fermi approximation without considering the HMT as one normally do in transport model simulations of heavy-ion collisions
are also shown. To illustrate the different HMT shapes considered in this work,  shown in Fig.~1 (b) are the momentum distributions of protons using $n(k) \sim1/k^4, 1/k^6$ and $1/k^9$, respectively. It is seen that the HMTs are different mainly at $k\ge 1.5$ fm$^{-1}$ (corresponding to about 300 MeV/c in momentum p).

In the IBUU04 transport model, an isospin- and momentum- dependent mean field single-nucleon potential
is used \cite{Das03}, i.e.,
\begin{eqnarray}
U(\rho,\delta,\textbf{p},\tau)&=&A_u(x)\frac{\rho_{\tau'}}{\rho_0}+A_l(x)\frac{\rho_{\tau}}{\rho_0}\nonumber\\
& &+B(\frac{\rho}{\rho_0})^{\sigma}(1-x\delta^2)
   -8x\tau\frac{B}{\sigma+1}\frac{\rho^{\sigma-1}}{\rho_0^\sigma}\delta\rho_{\tau'}\nonumber\\
& &+\frac{2C_{\tau,\tau}}{\rho_0}\int
d^3\,\textbf{p}'\frac{f_\tau(\textbf{r},\textbf{p}')}{1+(\textbf{p}-\textbf{p}')^2/\Lambda^2}\nonumber\\
& &+\frac{2C_{\tau,\tau'}}{\rho_0}\int
d^3\,\textbf{p}'\frac{f_{\tau'}(\textbf{r},\textbf{p}')}{1+(\textbf{p}-\textbf{p}')^2/\Lambda^2},
\end{eqnarray}
where $\tau=1/2\ (-1/2)$ for neutrons (protons),
$\delta=(\rho_n-\rho_p)/(\rho_n+\rho_p)$ is the isospin asymmetry,
and $\rho_n$, $\rho_p$ denote neutron and proton densities,
respectively. Considering the HMT, the parameters $A_u(x)$, $A_l(x)$, $B$,
$C_{\tau,\tau}$, $C_{\tau,\tau'}$ $\sigma$, and $\Lambda$ are readjusted
by fitting empirical properties of nuclear matte including the saturation density $\rho_0=0.16$ fm$^{-3}$, the binding energy $E_0=-16$ MeV,
the incompressibility $K_0=230$ MeV and the isoscalar effective mass
$M^*_s=0.7M$ where M is the average mass of nucleons in free space, see Ref.~\cite{Yong16} for more details. The $f_{\tau}(\textbf{r},\textbf{p})$ is
the phase-space distribution function of nucleons at coordinate \textbf{r} and
momentum \textbf{p}. Different $x$ parameters can be used to mimic
different density dependences of nuclear symmetry energy predicted by various nuclear many-body
theories without changing any property of symmetric nuclear matter
and the symmetry energy at normal density $\rho_0$ \cite{LCK08}. In this work, we choose
the symmetry energy parameter $x=1$ which is the original Gogny-Hartree-Fock prediction \cite{Das03}.
For nucleon-nucleon scattering cross sections, the isospin-dependent reduced in-medium nucleon-nucleon
cross sections \cite{LiChen05,Yong172,Gale02} are adopted.

One critical input for calculating hard photon productions perturbatively in heavy-ion reactions at intermediate energy is the
elementary cross section for the $p+n\rightarrow p+n+\gamma$ process \cite{Gan94,Nak86,Sch91,Timm06}.
This cross section is still somewhat model dependent. It is partially responsible for the fact that while the available experimental data
\cite{Gro86,Stevenson86,Tan88,Schu97} can be qualitatively well described by various transport models, some of the models mentioned earlier can reproduce quantitatively very well the data but others underpredicted the data by a factor as large as 5. Nevertheless, there is a strong consensus that the dominating mechanism for hard photon productions in heavy-ion collisions at intermediate energies is the incoherent neutron-proton bremsstrahlung \cite{Cass90}. Therefore, the focus of our exploratory work here is more on finding out qualitatively if any observables of hard photons may carry useful information about the shapes of the HMT rather than using existing data (actually not at high enough energy) to distinguish the shapes of the HMT (will be the goal of future experiments).
Since only relative effects of the different shapes of the HMT are important for the purposes of this study, the remaining uncertainties and model dependences of the elementary cross section for the $p+n\rightarrow p+n+\gamma$ process should not affect our conclusions qualitatively. We thus adopt the following probability of hard photon
production from the neutral scalar $\sigma$ meson exchange model \cite{Gan94}
\beq\label{pr}
\frac{d^2p_{\gamma}}{d\Omega dE_{\gamma}}=1.671\times10^{-7}\frac{(1-y^2)^{\alpha}}{y}
\eeq
where $y=E_{\gamma}/E_{\rm{max}}$, $\alpha=0.7319-0.5898\beta_i$, $E_{\gamma}$ is the energy of the emitted photon,
$E_{\rm{max}}$ is the total energy available in the proton-neutron center of mass (c.m.) system, while $\beta_i$
is the nucleon initial velocity. In the IBUU04 code, effects of the Pauli blocking in the final
state of the $p+n \rightarrow p+n+\gamma$ process are also taken into account as done in Ref. \cite{Bauer86}.
Moreover, photons are assumed to be emitted isotropically in the
proton-neutron c.m. frame. Therefore, one obtains the single differential elementary production probability $p_{\gamma}=\frac{dN}{dE_{\gamma}}$
by simply dividing the Eq. (\ref{pr}) by $4\pi$.

To avoid causing confusions, it is worth emphasizing several physical and technical points of our approach: (1) All nucleons including those in the HMT we distributed according to the momentum distribution of Eq. (\ref{mom}) are on-shell as guided by the derivations done using both the dilute interacting Fermi gas model, see, e.g., Eq. (5.24) in Ref. \cite{ABR} or the last Eq. on page 2 of Ref. \cite{Bel61} as well as the microscopic nuclear many-body theories, see, e.g., Eq. (43) of Ref. \cite{Rios13} within the Self-Consistent Greens Functions approach or the Eq. (11) in Ref. \cite{Yin} within the Extended Brueckner-Hartree-Fock approach. Such momentum distributions in nuclei have been explored using several different probes including electrons, photons, nucleons and nuclei and the results are consistent, see, e.g., examples given in Ref. \cite{Ant93}. 
(2) The photon production in heavy-ion collisions is calculated perturbatively by folding the time-dependent phase space distribution functions of single neutrons and protons given by the IBUU04 transport model with the elementary hard-photon production probability given above.  As argued in several earlier  papers, see, e.g., Ref. \cite{Niita}, since the hard photon production is so rare in heavy-ion reactions at Fermi energies and photons suffer no final state strong interaction, a perturbative approach is sufficient. As such, however,  the total energy of the whole reaction system is inherently not conserved as the nuclear reaction dynamics is not affected by the photon emission. (3) The IBUU04 transport model we use here has no cluster in the reaction process, there is thus no collision between nucleons and clusters. All photons are only from incoherent neutron-proton  
bremsstrahlungs. Moreover, the first collisions among nucleons in the same nucleus are forbidden by design to better satisfy the Pauli principle and keep the individual nucleus stable. Thus, essentially all hard photons are from collisions between protons (neutrons) from one nucleus with neutrons (protons) in the other nucleus. 
(4) The cascade model calculation in Ref. \cite{Ko85} indicates that while for low-energy photons ($\leq 30$ MeV) in reactions with heavy nuclei, there is a clear collective quadrupole bremsstrahlung, for higher energy photons or lighter nuclei, there is a large background coming from the incoherent dipole component of the bremsstrahlung. To our best knowledge, all available high-energy photon data ($\geq 50$ MeV) from intermediate energy heavy-ion reactions have been reasonably well explained using only the incoherent n+p bremsstrahlung, see, e.g., the review in Ref. \cite{Cass90}. The incoherent n+p bremsstrahlung is the mechanism we adopted in this work without considering any collective mechanism for the production of high energy photons. All of our results to be presented next should be understood within this context and with the cautions mentioned above.

%===============================================================================
\section{Results and Discussions}

\subsection{SRC effects on the production dynamics of hard photons}

\begin{figure*}[hbtp]%..............................................................
%\vspace{-4mm}
\centering
\setlength{\abovecaptionskip}{-5.cm}
\includegraphics[height=11cm,width=1.96\linewidth]{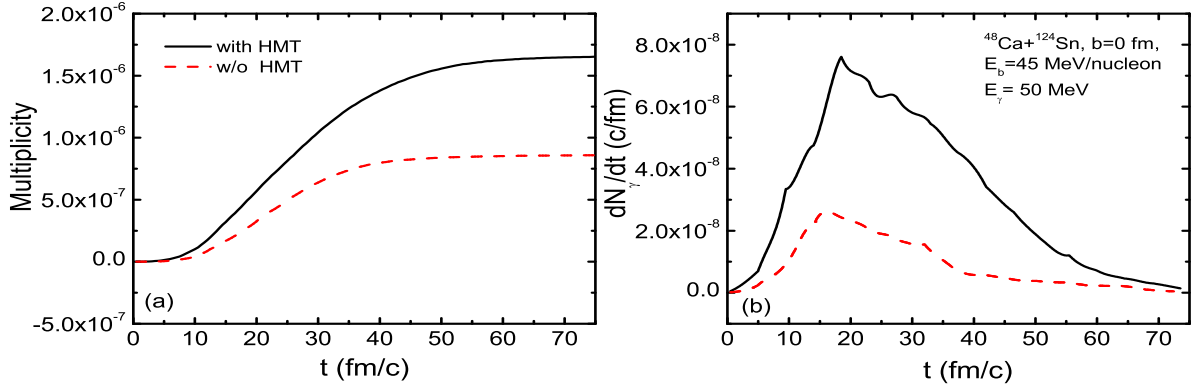}
\caption{Time evolutions of the multiplicity (a) and production rates (b) of hard photons with an energy of
$E_{\gamma}=50$ MeV in head-on collisions of $^{48}$Ca+$^{124}$Sn at a beam energy of 45 MeV/nucleon
with and without considering the HMT.} \label{N-dN}
\end{figure*}

\begin{figure}[t!]%..............................................................
\centering
\setlength{\abovecaptionskip}{-5.cm}
\includegraphics[width=1.\textwidth]{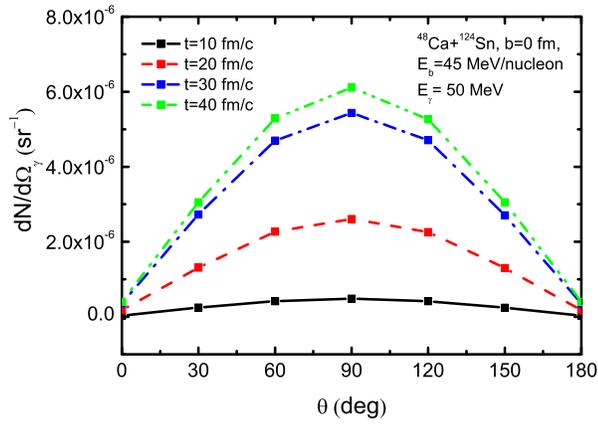}
\vspace{-1.2cm}
\caption{Angular distributions of hard photons of energy 50 MeV in the c.m. frame of the two colliding nuclei at
time $t=10,\ 20,\ 30,$ and $40$ fm/c in head-on $^{48}$Ca+$^{124}$Sn reactions at a beam energy of $E_b=45$ MeV/nucleon.} \label{Angular-t}
\end{figure}

\begin{figure}[t!]%..............................................................
\centering
\setlength{\abovecaptionskip}{-5.cm}
\includegraphics[width=1.\textwidth]{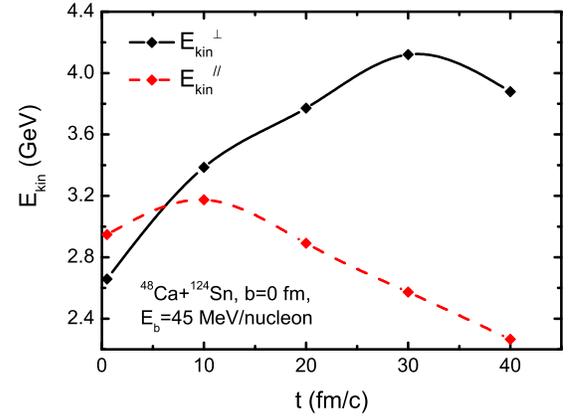}
\vspace{-1.2cm}
\caption{Time evolutions of transverse and longitudinal kinetic energy of nucleons in head-on $^{48}$Ca+$^{124}$Sn reactions at a beam energy of $E_b=45$ MeV/nucleon.} \label{Ekin-t}
\end{figure}

In Fig.~2, we show the time evolutions of the multiplicity (a) and production rate (b) of hard photons with an energy of
$E_{\gamma}=50$ MeV in head-on collisions of $^{48}$Ca+$^{124}$Sn at a beam energy of 45 MeV/nucleon
with and without considering the HMT. Here, the $1/k^4$ shape of HMT is used. It is clearly seen that the HMT leads to an obvious increase of both the production rate and final multiplicity of hard photons.
Without the HMT, the production of hard photons stopped and the multiplicity saturated after about 40 fm/c. With the HMT, however, the production lasts significantly longer and produce more hard photons. This is what one expects as not only collisions between two HMT nucleons from the target and projectile, but also collisions involving one HMT nucleon and one nucleon below the Fermi surface in either one of the two colliding nuclei may be energetic enough to produce high energy photons. It also takes more collisions to slow down HMT nucleons, therefore not only the first chance but also secondary or third collisions involving  HMT nucleons may contribute to the production of high energy photons.

Fig.~3 shows the angular distributions of hard photons with energy $E_{\gamma}=50$ MeV in the c.m. frame of the colliding nuclei at
time $t=10,\ 20,\ 30,$ and $40$ fm/c in head-on collisions of $^{48}$Ca+$^{124}$Sn at a beam energy of 45 MeV/nucleon.  It is interesting to see that while at the earlier stage of the reaction around t=10 fm/c, the photon angular distribution is almost isotropic, it gradually peaks around $\theta=90^{\circ}$. This is due to the reaction dynamics. As mentioned earlier, the elementary photon production cross section is isotropic in the neutron-proton c.m. frame.
The first chance neutron-proton collisions are mostly those involving one nucleon from the target and another from the projectile. Their centers of mass are approximately at rest in the c.m. frame of the colliding nuclei (similar to SRC neutron-proton pairs having a small c.m momentum but a large relative momentum), therefore the resulting photon angular distribution is approximately isotropic in the c.m. frame of the colliding nuclei. However, as the reaction evolves
the nucleon momentum distribution gets modified in a way that nucleons (thus neutron-proton c.m. momenta) move preferentially vertical to the beam direction due to the well-known transverse flow and/or squeeze-out phenomena.
Since some of these predominantly vertical moving nucleons may be energetic enough to produce hard photons (thus, the photon source directions), hard photons develop a peak around $\theta=90^{\circ}$. This picture is more
clearly seen by examining the evolution of the nucleon kinetic energies in the longitudinal and transverse directions in Fig.~4. It is seen that the total transverse kinetic energy of nucleons increases continuously up to bout 30 fm/c, while the longitudinal kinetic energy first increases for a short time due to the initial attractive nuclear force then decreases continuously as energies are being transferred to the vertical direction. This is a reflection of the nuclear stopping process and the development of some collectivity in the transverse direction. Consequently, collisions between nucleons moving in the vertical direction increase gradually, leading eventually to more photon emissions around 90 degrees in the c.m. frame of the two colliding nuclei.

\subsection{Impact parameter and beam energy dependence of hard photon productions}

\begin{figure}[t!]%..............................................................
\centering
\setlength{\abovecaptionskip}{-5.cm}
\includegraphics[width=1.\textwidth]{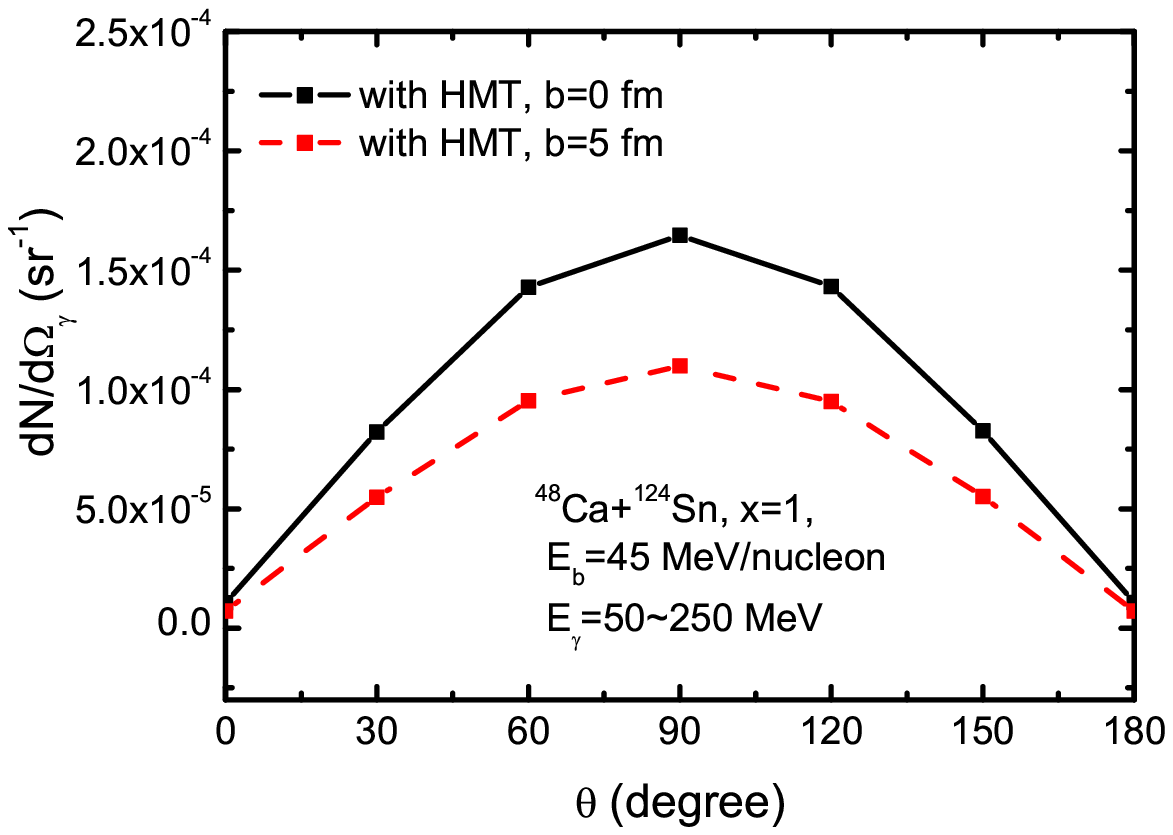}
\vspace{-1cm}
\caption{Comparison of angular distributions of hard photons with energies between 50 and 250 MeV in the c.m. frame of the colliding nuclei
in the head-on ($b=0$ fm)
and mid-central ($b=5$ fm) collisions of $^{48}$Ca+$^{124}$Sn with HMTs ($\sim 1/k^4$).} \label{Ang-hmts}
\end{figure}

\begin{figure}[t!]%..............................................................
\centering
\setlength{\abovecaptionskip}{-5.cm}
\includegraphics[width=1.\textwidth]{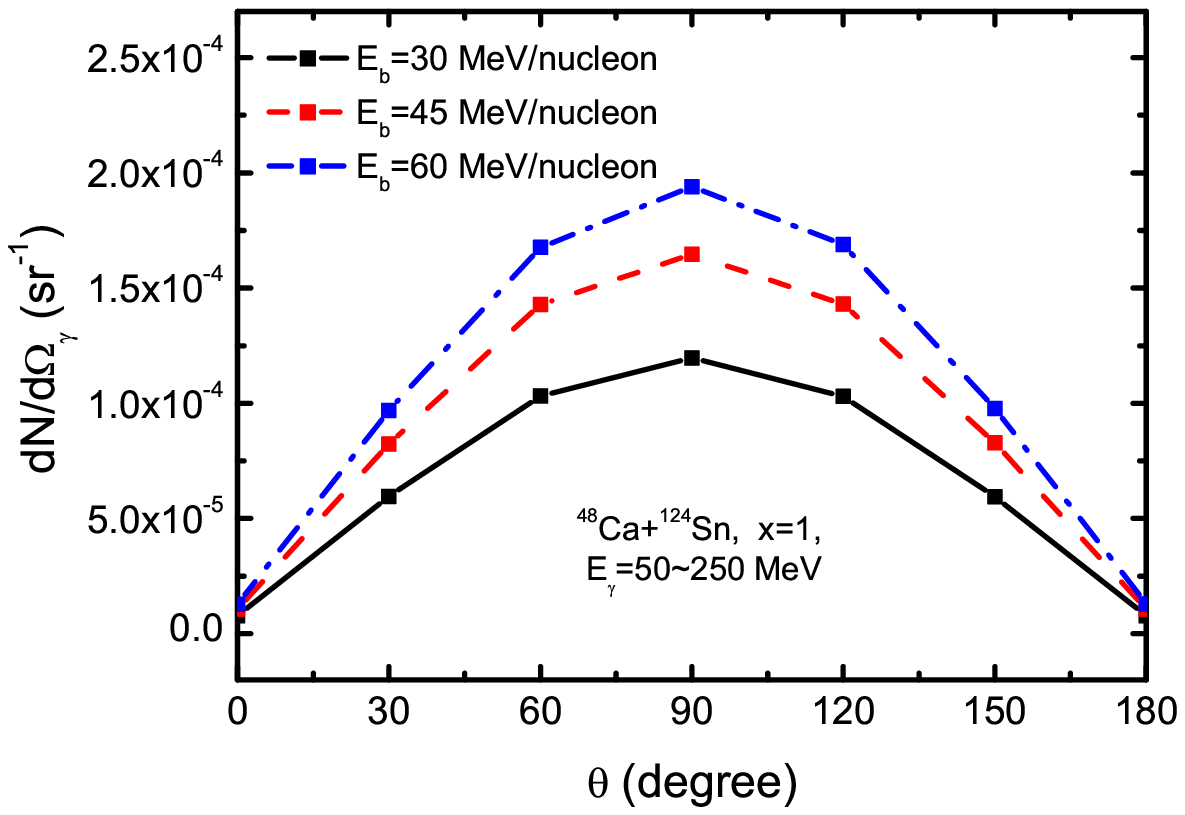}
\vspace{-1cm}
\caption{Angular distributions of hard photons with energies between 50 and 250 MeV
in the c.m. frame of the colliding nuclei with HMTs  ($\sim 1/k^4$) for head-on $^{48}$Ca+$^{124}$Sn reactions at beam
energies of 30, 45, and 60 MeV/nucleon, respectively.}
\end{figure}

In order to understand completely the dynamics of hard photon productions in heavy-ion collisions, it is necessary to study systematically the impact parameter and beam energy dependences of hard photon productions.
The results should be particularly useful for planning future experiments \cite{Alan}. For this purpose, shown in Fig.~5 is a comparison of the single differential probability of hard photons with energies between 50 and 250 MeV in
head-on ($b=0$ fm) and mid-central ($b=5$ fm) collisions of $^{48}$Ca+$^{124}$Sn at a beam energy of $E_{\rm{b}}=45$ MeV/nucleon with HMTs ($\sim 1/k^4$).
It is seen that the single differential probability $dN/d\Omega_{\gamma}$ in the head-on collision is lager by
a factor of about 1.5 at $\theta=90^{\circ}$ compared to the mid-central reaction with an impact parameter of $b=5$ fm.
This is simply because of the more abundant proton-neutron collisions in the head-on reaction.

Shown in Fig.~6 is the beam energy dependence of hard photon productions.
Here, the angular distributions of the single differential probability of hard photons with energies between 50 and 250 MeV
in $^{48}$Ca+$^{124}$Sn reactions with HMTs ($\sim 1/k^4$) at beam energies of
30, 45, and 60 MeV/nucleon are compared.
As expected, more photons are produced in reactions at higher beam energies over the whole angular
range considered. More quantitatively, the single differential probability at $E_{\rm{b}}=60$
MeV/nucleon increases by a factor of about 1.2 compared to $E_{\rm{b}}=45$ MeV/nucleon,
and 1.6 to $E_{\rm{b}}=30$ MeV/nucleon at $\theta=90^{\circ}$. Apparently, the higher beam
energy makes the colliding nucleons more energetic, leading to more emissions
of hard photons from the $n+p\rightarrow n+p+\gamma$ processes.

\subsection{Gauging the model with the old data}
\begin{figure}[t!]%..............................................................
\centering
\setlength{\abovecaptionskip}{-5.cm}
\includegraphics[width=1.\textwidth]{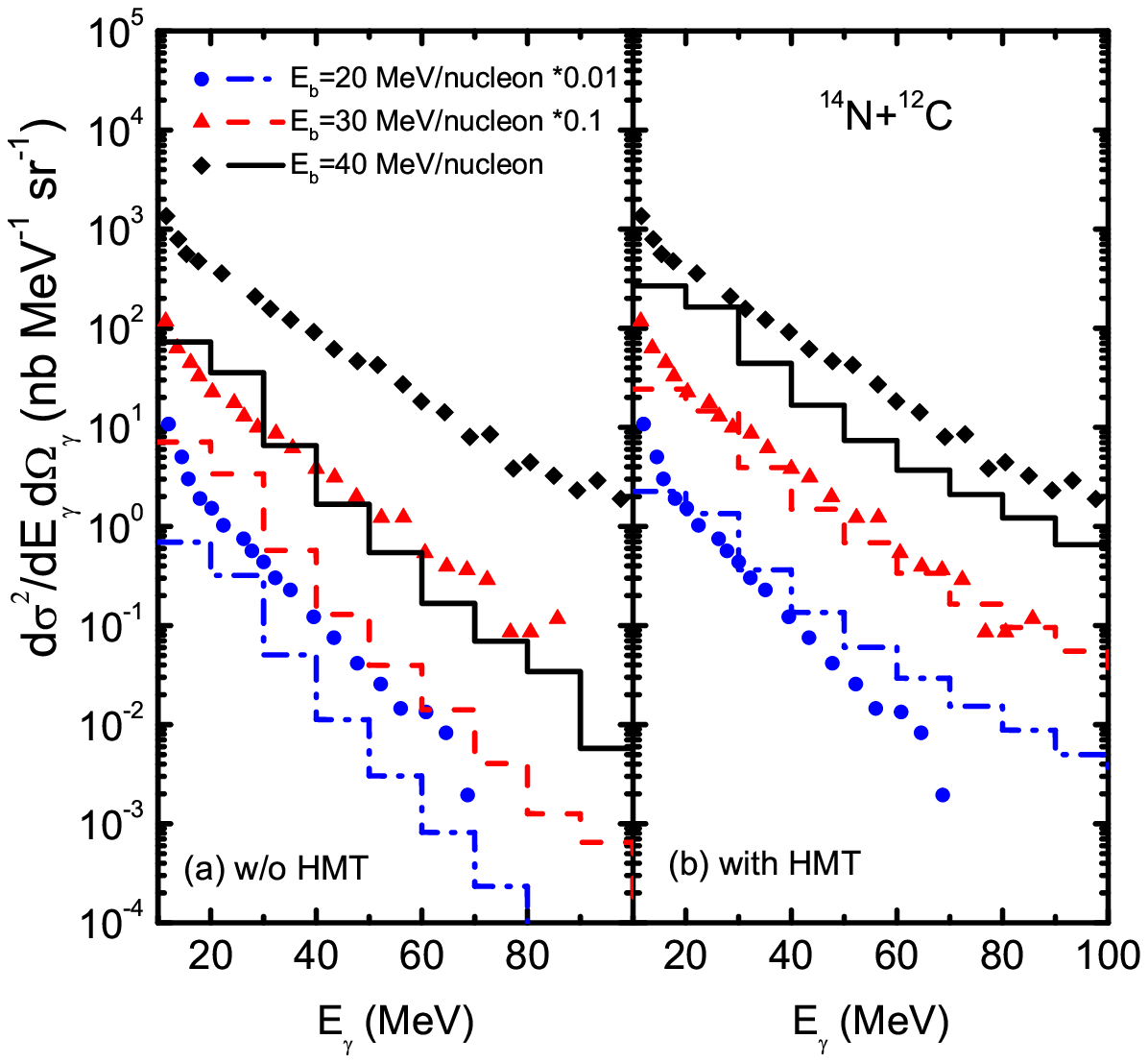}
\vspace{-3.5cm}
\caption{Photon energy spectra from $^{14}$N+$^{12}$C reactions at a beam energy of $E_b=20$ (blue), 30 (red) and 40 (black) MeV/nucleon, respectively. The step lines in panel (a) are the results without the HMT,
while the step lines in panel (b) are results with the HMT ($\sim 1/k^4$) in the IBUU04 model. The solid symbols represent the experimental data taken from Ref.~\cite{Stevenson86}. } \label{exp1}
\end{figure}

\begin{figure}[t!]%..............................................................
\centering
\setlength{\abovecaptionskip}{-7. cm}
\setlength{\belowcaptionskip}{-0.53 cm}
\includegraphics[width=1.\textwidth]{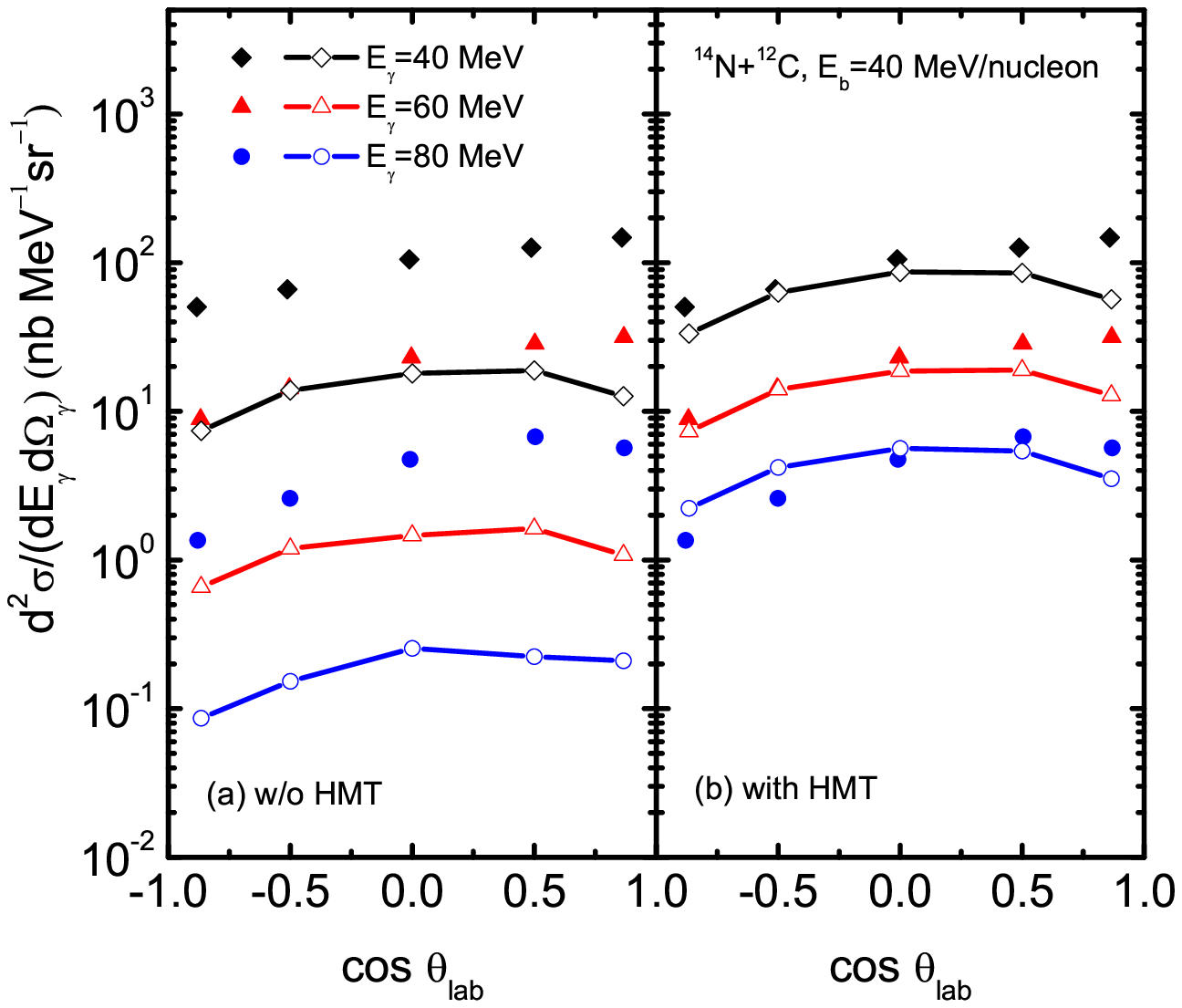}
\vspace{-1.2cm}
\caption{Angular distributions of photons with energies of $40\pm3$ (black), $60\pm3$ (red), and $80\pm3$ (blue) MeV, respectively, from the minimum bias $^{14}$N+$^{12}$C reactions at a beam energy of $E_b=40$ MeV/nucleon in the laboratory frame. The lines with open markers in panel (a) are the results without considering the HMT, while the lines with open markers in panel (b) are the results with HMTs in the IBUU04 model. The solid symbols represent the experimental data taken from Ref.~\cite{Stevenson86}. } \label{exp2}
\end{figure}

As mentioned earlier, there were some interesting data of hard photons from intermediate energy heavy-ion reactions in the earlier 80's \cite{Gro86,Stevenson86,Tan88,Schu97}. Although not all the data are fully compatible, they stimulated many theoretical studies. Generally, the data can be understood qualitatively as due to mainly $p+n\rightarrow p+n+\gamma$ processes. The level of quantitative agreements between model calculations and the experimental data has been model dependent. While some models can well reproduce the data, others miss significantly both the multiplicities and the angular distributions of hard photons.

As indicated earlier, while our main goal is not to re-analyze quantitatively the old data but to study more qualitatively if the shapes of the HMT leave any imprint on any observable of hard photons, it is interesting and useful to get an orientation by comparing our model calculations with some of the old experimental data.  For this purpose, shown in Fig.~7 are comparisons of the calculated energy spectra of hard photons and the experimental data
at $\theta_{lab}\approx 90^{\circ}$ from Ref.~\cite{Stevenson86} for minimum bias $^{14}$N+$^{12}$C reactions at a beam energy of $E_b=20$ (blue), 30 (red) and 40 (black) MeV/nucleon, respectively.
The results calculated without considering the HMT are plotted in panel (a) while those with the HMTs ($\sim 1/k^4$) in panel (b).
The experimental energy spectrum shows an exponential fall-off with increasing hard photon energies. Without the HMT, the calculations underpredict significantly the yields of hard photons at all beam energies considered. Apparently, with the HMT, the calculations can better describe the data. While the calculations with the HMT can not quantitatively reproduce all spectra in the whole photon energy range,
overall, the calculations with the HMT are much more closer to the data compared to the ones without considering the HMT, indicating that the latter plays a significant role.

Similar information about the importance of HMT can be obtained from comparing the calculated and measured angular distributions.
Fig.~8 shows the angular distributions of hard photons with energies of
$40\pm3$ (black), $60\pm3$ (red), and $80\pm3$ (blue) MeV, respectively, from the  minimum bias $^{14}$N+$^{12}$C reaction at a beam energy of $E_b=40$ MeV/nucleon in the laboratory frame.
The lines with open markers in panel (a) are the results without the HMT,
while the lines with open markers in panel (b) are the results with HMTs in the IBUU04 model.
The solid symbols represent the experimental data taken from Ref.~\cite{Stevenson86}.
It is seen that the angular distributions are all slightly forward peaked. This is because for this approximately symmetric reaction, the c.m. of the colliding nuclei is moving forward with a half of the beam velocity.
While the photon spectra peak around $90^{\circ}$ in the c.m. of the colliding nuclei, they become slightly forward peaked in the laboratory frame.
Comparing Fig.~8 (a) and (b), we can also see that our calculations are in better agreement
with the experimental data when the HMTs are considered.
Again, it indicates that the HMT plays an important role in hard photon production in intermediate energy
heavy-ion collisions. Hence, the energy spectra and angular distributions of hard photons may carry some useful information about the shape of the HMTs in nucleon momentum distributions in the colliding nuclei.
All indications from the above comparisons thus further call for a study about effects of different shapes of HMT on the production of hard photons.

\subsection{Probing the shape of HMT with the angular distribution, energy and transverse momentum spectra of hard photons}
\begin{figure}[hbtp]%..............................................................
\centering
\setlength{\abovecaptionskip}{-6.cm}
\includegraphics[width=1.\textwidth]{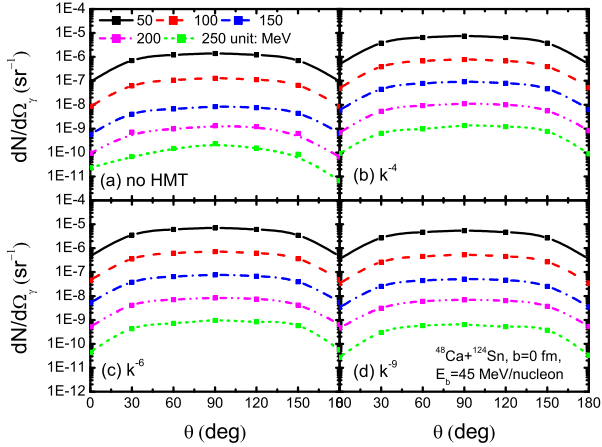}
\vspace{-1.2cm}
\caption{Angular distributions of photons with energies of $E_{\gamma}=50,\ 100,\ 150,\ 200,$ and $250$ MeV in the c.m. frame of the colliding nuclei from
the head-on $^{48}$Ca+$^{124}$Sn reaction at a beam energy of $E_b=45$ MeV/nucleon with different HMT shapes.} \label{double}
\end{figure}

\begin{figure}[hbtp]%..............................................................
\centering
\includegraphics[height=6.9cm,width=9cm]{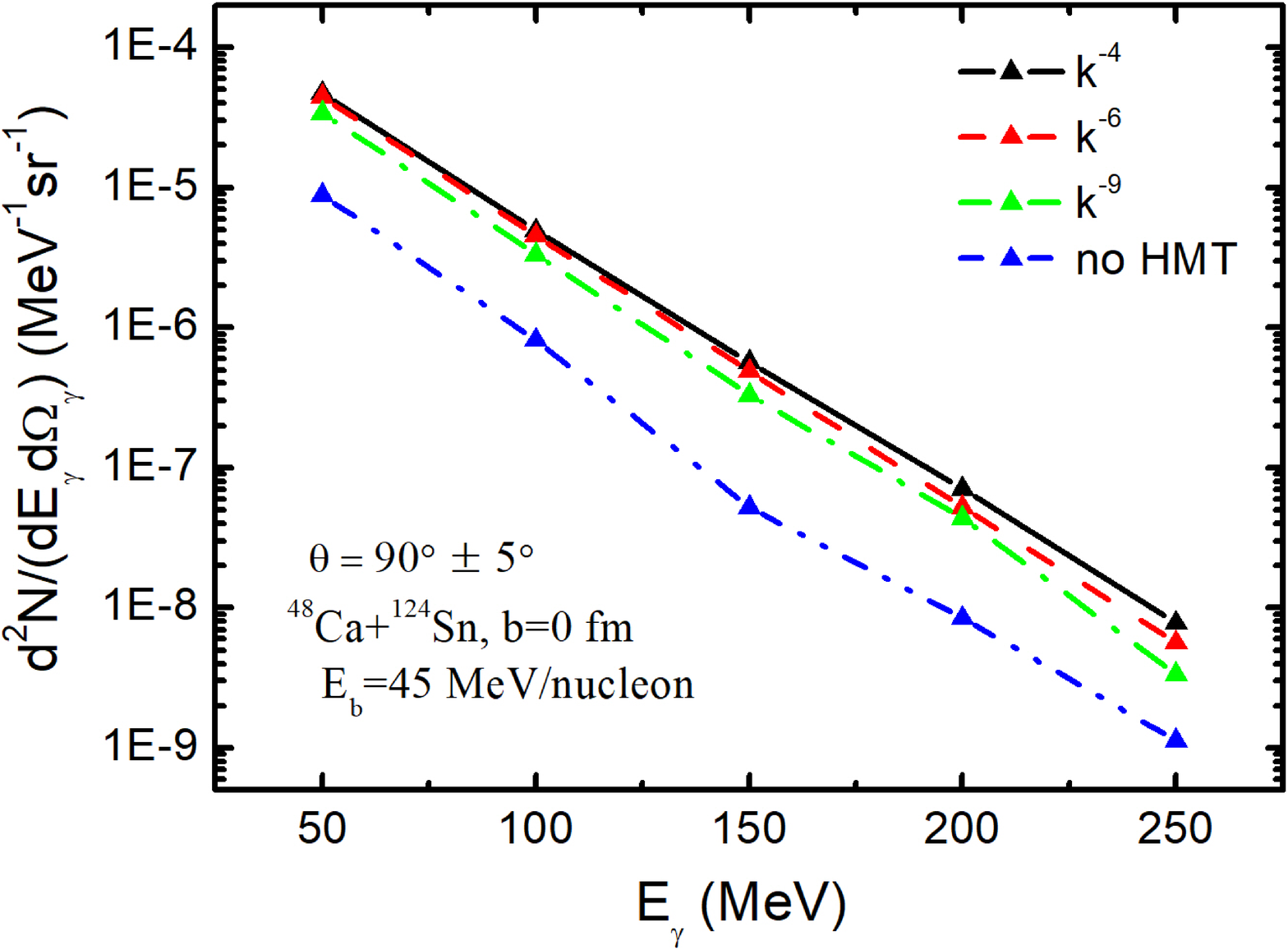}
\caption{Effects of different shapes of HMT on the energy spectrum of hard photons emitted within the polar angular range of $85^{\circ}\leq \theta \leq 95^{\circ}$ in the c.m frame of the head-on $^{48}$Ca+$^{124}$Sn
reaction with a beam energy of 45 MeV/nucleon. } \label{E-hmts}
\end{figure}

\begin{figure}[t!]%..............................................................
\centering
\setlength{\abovecaptionskip}{-5.cm}
\includegraphics[width=1.\textwidth]{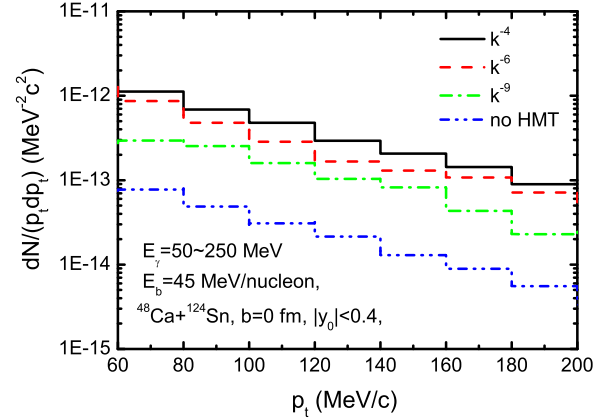}
\vspace{-1cm}
\caption{The transverse momentum dependence of hard photon spectra in the mid-rapidity region of $-0.4\leq y_0\leq 0.4$ in the head-on reaction of
$^{48}$Ca+$^{124}$Sn at a beam energy of $E_b=45$ MeV/nucleon with different shapes of HMT.} \label{pt}
\end{figure}

We now turn to examine effects of different shapes of HMT on hard photon production. Shown in Fig.~9 are the angular distributions of photons with energies of $E_{\gamma}=50,\ 100,\ 150,\ 200$ and $250$ MeV in the c.m. frame of the colliding nuclei from the head-on $^{48}$Ca+$^{124}$Sn reaction at a beam energy of $E_b=45$ MeV/nucleon with different HMT shapes. Compared to the results without considering the HMT, while the yields of hard photons obviously increases with all three shapes $n(k) \sim1/k^4, 1/k^6$ and $1/k^9$ as one expects, the symmetric feature of the angular distribution remains the same. This is understandable as in our model the angular distribution depends only on the energy of the colliding neutron-proton pairs according to the Eq. (\ref{pr}). The shape of HMT has no effect on the angular distributions of the produced photons. The yields of high energy photons decreases exponentially with increasing photon energy as expected and this feature is consistent with the findings in Refs.~\cite{Bauer86,Ko87,Yong17}. Therefore, we can conclude that the angular distributions of hard photons will not help probe the shapes of HMTs in the colliding nuclei.

Next we examine the energy spectra of hard photons. Since hard photons are most abundant around $\theta=90^{\circ}$ in the c.m. frame of colliding nuclei as shown above,
we thus further study the energy spectra of these hard photons near $90^{\circ}$.
Shown in Fig.~10 are the energy spectra of hard photons emitted within the polar angle range of $85^{\circ}\leq \theta \leq 95^{\circ}$ in the head-on $^{48}$Ca+$^{124}$Sn reactions at a beam energy of 45 MeV/nucleon.
Effects of different shapes of HMT on the hard photon energy spectra are appreciable. Moreover, there is a clear indication that the HMT shape effect becomes more evident for more energetic photons.
This is because photons with less energies originate mainly from collisions of nucleons
from either one HMT and one Fermi sea or two Fermi seas. While the more energetic hard photons are from collisions of two nucleons both from the HMTs of the target and projectile.
They are thus more sensitive to the shapes of the HMT.

As discussed earlier, the development of the peak around $\theta=90^{\circ}$ in the photon angular distribution in the c.m. frame of colliding nuclei is associated with the transfer of nucleon momentum from the beam direction to the transverse direction as the reaction proceeds. It is thus interesting to examine how the hard photon transverse momentum distribution may depend on the shape of the HMT. Shown in Fig.~11 are the transverse momentum distributions of the hard photons spectra, $dN/p_tdp_t$ versus $p_t$ ($p_t=\sqrt{p^2_x+p^2_y}$)  in the mid-rapidity region of $-0.4\leq y_0\leq 0.4$
($y_0=y/y_{\rm{beam}}$ is the ratio of particle rapidity y over beam rapidity $y_{\rm{beam}}$ in the c.m. frame of the colliding nuclei), for head-on collisions of $^{48}$Ca+$^{124}$Sn at a beam energy of $E_{\rm{b}}=45$ MeV/nucleon with different shapes of the nucleon HMT. It is seen that the spectra show typical exponential shapes for $p_t\geq 50$ MeV/c with all three shapes of the HMT considered.
Interestingly,  effects of the HMT on the transverse momentum spectrum are more obvious compared to the HMT effects on the energy spectrum shown in Fig.~10.
Moreover, the shape of the HMT has a more clear imprint on the transverse momentum spectrum of hard photons, making the latter a more useful observable for investigating the shape of the nucleon HMT.

\section{Summary}
In summary, motivated by the impressive recent progress in exploring the nature of nucleon SRC in nuclei using electron/nucleon-nucleus scatterings and the emerging interest to further explore the shape of the nucleon HMT caused by the SRC, we studied effects of nucleon HMT on hard photon productions in heavy-ion reactions at beam energies around the Fermi energy. Our main goal is to look for imprints of the different shapes of the HMT on experimental observables of hard photons. The shape of the nucleon HMT is directly related to the poorly known properties of the short-range nuclear forces. It has important ramifications in both nuclear physics and astrophysics.

Within the IBUU04 transport model incorporating approximately the nucleon HMT induced by the SRC in colliding nuclei, we investigated hard photon emissions in $^{14}$N+$^{12}$C and $^{48}$Ca+$^{124}$Sn reactions. Effects of different shapes of the HMT on the energy spectrum, angular distribution and transverse momentum spectrum of hard photons were studied.  While the angular distribution does not carry any information about the shape of the HMT, the energy spectra and especially the mid-rapidity transverse momentum spectra of hard photons are found to bare strong imprints of the shapes of HMTs in the two colliding nuclei. This information is expected to be useful for planning future experiments. We notice that the yields of very high energy photons from heavy-ion collisions around the Fermi energy are very low. Thus, the experiments might be very challenging, but we assume they are not more difficult than measuring gravitational waves from occasional collisions of two neutron stars somewhere in space.

\begin{acknowledgments}
We would like to thank Alan B. McIntosh and Giuseppe Verde for helpful communications, Bao-Jun Cai and Wei Zuo for useful discussions. The work is partially supported by the National Natural Science Foundation of China under Grant Nos.~11705109, 11775275, the Key Program of Innovation and Entrepreneurship Support Plan for Returned Talents in Anhui Province under Grant No.~2020LCX011,
and the Institute of energy, Hefei Comprehensive National Science Center under Grant No.~GXXT-2020-004.
B. A. Li is supported in part by the U.S. Department of Energy, Office of Science, under Award Number DE-SC0013702, the CUSTIPEN (China-U.S. Theory Institute for Physics with Exotic Nuclei) under the US Department of Energy Grant No. DE-SC0009971.

\end{acknowledgments}

%===============================================================================


\begin{thebibliography}{99}

\bibitem{Mig57} A. B. Migdal, Sov. Phys. JEPT. \textbf{5}, 333 (1957).

\bibitem{bethe} H. A. Bethe, Ann. Rev. Nucl. Part. Sci. \textbf{21}, 93 (1971).

\bibitem{pan92} V. R. Pandharipande and S. C. Pieper, Phys. Rev. C \textbf{45}, 791 (1992).

\bibitem{Pan99} V. R. Pandharipande, I. Sick, and P. K. A. deWitt
Huberts, Rev. Mod. Phys. \textbf{69}, 981 (1999).

\bibitem{Ben93}O. Benhar, V. R. Pandharipande and Steven C. Pieper,
Rev. Mod. Phys. \textbf{93}, 817 (1993).

\bibitem{Gal58} V. M. Galitskii, Zh. Eksp, Teor. Fiz. \textbf{34}, 151 (1958).
\newblock [Sov. Phys. JETP \textbf{7}, 104 (1958)].

\bibitem{Lut60} J. M. Luttinger, Phys. Rev. \textbf{119}, 1153 (1960).
\bibitem{Czy61} W. Czyz and K. Gottfried, Nucl. Phys. A \textbf{21}, 676 (1961).
\bibitem{Bel61} V. A. Belykov, Sov. Phys. JETP. \textbf{13}, 850 (1961).
\bibitem{Sar80} R. Sartor and C. Mahaux, Phys. Rev. C \textbf{21}, 1546 (1980).

\bibitem{Sar82} R. Sartor and C. Mahaux, Phys. Rev. C \textbf{25}, 677 (1982).

\bibitem{Amado76a} R. D. Amado and R. M. Woloshyn, Phys. Lett. B \textbf{62}, 253
(1976).

\bibitem{Amado76b} R. D. Amado,  Phys. Rev. C \textbf{14}, 1264 (1976).

\bibitem{Amado77} R. D. Amado and R. M. Woloshyn,  Phys. Rev. C \textbf{15}, 2200
(1977).

\bibitem{Ant93}A. N. Antonov, P. E. Hodgson, and I. Zh. Petkov, Nucleon Correlations in Nuclei, Springer-Verlag, Berlin, Heidelberg 1993, ISBN 3-540-55911-6.

\bibitem{Hen16x}O. Hen, G. A. Miller, E. Piasetzky, and L. B. Weistein, Rev. Mod. Phys. \textbf{89}, 045002 (2017).

\bibitem{Ohen15}
O. Hen, B. A. Li, W. J. Guo, L. B. Weinstein, E. Piasetzky, Phys. Rev. C \textbf{91}, 044610 (2015).


\bibitem{PPNP} B. A. Li, B. J. Cai, L. W. Chen, and J. Xu, Prog. Part. Nucl. Phys. \textbf{99}, 29 (2018).

\bibitem{Cai15}
B. J. Cai and B. A. Li,
%``Symmetry energy of cold nucleonic matter within a relativistic mean field model encapsulating effects of high momentum nucleons induced by short-range correlations,''
Phys. Rev. C \textbf{93}, 014619 (2016).

\bibitem{Souza1}
L. A. Souza, R. Negreiros, M. Dutra, D. P. Menezes and O. Louren\c{c}o,
%``Short-range correlation effects on the neutron star cooling,''
[arXiv: 2004.10309 [nucl-th]].

\bibitem{Souza20}
L. A. Souza, M. Dutra, C. H. Lenzi and O. Louren\c{c}o,
%``Effects of short-range nuclear correlations on the deformability of neutron stars,''
Phys. Rev. C \textbf{101}, 065202 (2020).

\bibitem{Lu21}
H. Lu, Z. Ren and D. Bai,
%``Impacts of nucleon-nucleon short-range correlations on neutron stars,''
Nucl. Phys. A \textbf{1011}, 122200 (2021).


\bibitem{Hen15} O. Hen, L. B. Weinstein, E. Piasetzky, G. A. Miller, M. Sargsian, and Y. Sagi,
Phys. Rev. C \textbf{92}, 045205 (2015).

\bibitem{Rios13}
A. Rios, A. Polls and W. H. Dickhoff,
%``Density and isospin asymmetry dependence of high-momentum components,''
Phys. Rev. C \textbf{89}, 044303 (2014).


\bibitem{Ant07} A. N. Antonov, M. V. Ivanov, M. K. Gaidarov, E. Moya de Guerra, Phys. Rev. C \textbf{75}, 034319 (2007).

\bibitem{Li00}
B. A. Li,
%``Uranium on uranium collisions at relativistic energies,''
Phys. Rev. C \textbf{61}, 021903 (2000).

\bibitem{HLi20}
H. Li, H. j. Xu, Y. Zhou, X. Wang, J. Zhao, L. W. Chen and F. Wang,
%``Probing the neutron skin with ultrarelativistic isobaric collisions,''
Phys. Rev. Lett. \textbf{125}, 222301 (2020).

\bibitem{Jia21}
G. Giacalone, J. Jia and C. Zhang,
%``The impact of nuclear deformation on relativistic heavy-ion collisions: assessing consistency in nuclear physics across energy scales,''
[arXiv: 2105.01638 [nucl-th]].

\bibitem{Gia20}
G. Giacalone,
%``Constraining the quadrupole deformation of atomic nuclei with relativistic nuclear collisions,''
Phys. Rev. C \textbf{102}, 024901 (2020).

\bibitem{Ma18}
X. G. Deng and Y. G. Ma, Eur. Phys. J. A \textbf{54}, 204 (2018).

\bibitem{Giul06}
G. Giuliani, M. Papa, Phys. Rev. C \textbf{73}, 031601(R) (2006).

\bibitem{Bertsch88}
G. F. Bertsch and S. Das Gupta, Phys. Rep. \textbf{160}, 189 (1988).

\bibitem{Cass90}
W. Cassing,  V Metag, U Mosel and K Niita, 
Phys. Rep. \textbf{188}, 363 (1990).

\bibitem{Bona94}
A.~Bonasera, F.~Gulminelli and J.~Molitoris,
%``The Boltzmann equation at the borderline: a decade of Monte Carlo simulations of a quantum kinetic equation,''
Phys. Rept. \textbf{243}, 1 (1994).

\bibitem{Bona06}
A. Bonasera, R. Coniglione, and P. Sapienza, Eur. Phys. J. A \textbf{30}, 47 (2006).

\bibitem{Ko85}
C. M. Ko, G. Bertsch and J. Aichelin,
%``PROBING HEAVY ION COLLISIONS WITH BREMSSTRAHLUNG,''
Phys. Rev. C \textbf{31}, 2324 (1985).

\bibitem{Cassing86}
W. Cassing, T. Biro, U. Mosel, M. Tohyama and W. Bauer,
%``High energy \ensuremath{\gamma}-rays: A probe for momentum- and energy-distributions in the reaction zone?,''
Phys. Lett. B \textbf{181}, 217 (1986).

\bibitem{Nak86}
K. Nakayama and G. F. Bertsch,
%``High energy photon production in nuclear collisions,''
Phys. Rev. C \textbf{34}, 2190 (1986).

\bibitem{Bauer86}
W. Bauer, G. F. Bertsch, W. Cassing, and U. Mosel, Phys. Rev. C \textbf{34}, 2127 (1986).

\bibitem{Bauer86a}
W. Bauer, W. Cassing, U. Mosel, M. Tohyama and R. Y. Cusson,
%``High energy \ensuremath{\gamma}-ray emission in heavy-ion collisions,''
Nucl. Phys. A \textbf{456}, 159 (1986).

\bibitem{Rem86}
B. A. Remington, M. Blann and G. F. Bertsch,
%``Intranuclear N-N Collision Model for the Production of High-Energy Gamma Rays in Heavy-Ion Collisions,''
Phys. Rev. Lett. \textbf{57}, 2909 (1986).

\bibitem{Ko87}
C. M. Ko, J. Aichelin, Phys. Rev. C \textbf{35}, 1976 (1987).

\bibitem{Biro87}
T. Biro, K. Niita, A. D. Paoli, W. Bauer, W. Cassing, and U. Mosel, Nucl. Phys. A
\textbf{475}, 579 (1987).

\bibitem{Niita}
K.~Niita, A.~L.~De Paoli, W.~Bauer, T.~S.~Biro, W.~Cassing and U.~Mosel,
%``Hard protons from heavy-ion collisions,''
Nucl. Phys. A \textbf{482}, 525 (1988).

\bibitem{Gan94}
N. Gan, K.-T. Brinkmann, A. L. Caraley, et al., Phys. Rev. C \textbf{49}, 1 (1994).

\bibitem{Liu08}
G. H. Liu, Y. G. Ma, X. Z. Cai, D. Q. Fang, W. Q. Shen, W. D. Tian, and K. Wang,
Phys. Lett. B \textbf{663}, 312 (2008).

\bibitem{Yong08}
G. C. Yong, B. A. Li, and L. W. Chen, Phy. Lett. B \textbf{661}, 82 (2008).

\bibitem{Shicz20}
C. Z. Shi, Y. G. Ma, X. G. Cao, D. Q. Fang, W. B. He, and C. Zhong,
Phys. Rev. C \textbf{102}, 014601 (2020).

\bibitem{WangSS20}
S. S. Wang, Y. G. Ma, X. G. Cao, D. Q. Fang, and C. W. Wang,
Phys. Rev. C \textbf{102}, 024620 (2020).

\bibitem{Yong17}
G. C. Yong, and B. A. Li, Phys. Rev. C \textbf{96}, 064614 (2017).
\bibitem{Yong172}
G. C. Yong, Phys. Lett. B \textbf{765}, 104 (2017).
\bibitem{Yong18}
G. C. Yong, Phys. Lett. B \textbf{776}, 447 (2018).

\bibitem{Alan} Alan B. McIntosh, private communications.

\bibitem{Giu} Giuseppe Verde, private communications.

\bibitem{LiBA04a} B. A. Li, C. B. Das, S. Das Gupta and C. Gale, Phys. Rev. C \textbf{69}, 011603 (2004).

\bibitem{LiBA04b} B. A. Li, C. B. Das, S. Das Gupta and C. Gale, Nucl. Phys. A \textbf{735}, 563 (2004).

\bibitem{LiChen05} B. A. Li and L. W. Chen, Phys. Rev. C \textbf{72}, 064611 (2005).

\bibitem{Liba05}B. A. Li, G. C. Yong, and W. Zuo, Phys. Rev. C \textbf{71}, 014608 (2005).

\bibitem{Frie86}
J. Friedrich and P. G. Reinhard, Phys. Rev. C \textbf{33}, 335 (1986).

\bibitem{Cai16}
B. J. Cai, B. A. Li, Phys. Rev. C \textbf{93}, 014619 (2016).

\bibitem{Sube08}
R. Subedi, et al., Hall A. Collaboration, Science \textbf{320}, 1476 (2008).

\bibitem{OHen14}
O. Hen, et al. (The CLAS Collabration), Science \textbf{346}, 614 (2014).

\bibitem{Das03}
C. B. Das, S. Das Gupta, C. Gale, and B. A. Li, Phys. Rev. C \textbf{67}, 034611 (2003).

\bibitem{Yong16}
G. C. Yong, Phys. Rev. C \textbf{93}, 044610 (2016).

\bibitem{LCK08} B. A. Li, L. W. Chen, and C. M. Ko, Phys. Rep. \textbf{464}, 113 (2008).

\bibitem{Gale02} D. Persram and C. Gale, Phys. Rev. C \textbf{65}, 064611 (2002).

\bibitem{Sch91}
M. Sch\"{a}fer, T. S. Biro, W. Cassing, U. Mosel, H. Nifenecker,
and J. A. Pinstan, Z. Phys. A \textbf{339}, 391 (1991).

\bibitem{Timm06}
R. G. E. Timmermans, T. D. Penninga, B. F. Gibson, and M. K. Liou,
Phys. Rev. C \textbf{73}, 034006 (2006).

\bibitem{Gro86}
E. Grosse, P. Grimm, H. Heckwolf, F. J. Mueller, H. Noll,
A. Oskarson, H. Stelzer, and W. Roesch, Europhys. Lett. \textbf{2}, 9
(1986).

\bibitem{Stevenson86}
J. Stevenson, K. B. Beard, W. Benenson, J. Clayton, E. Kashy, A. Lampis,
D. J. Morrissey, M. Samuel, R. J. Smith, C. L. Tam, and J. S. Winfield,
Phys. Rev. Lett. \textbf{57}, 555 (1986).

\bibitem{Tan88}
C. L. Tam, J. Stevenson, W. Benenson, J. Clayton, Y. Chen, E. Kashy,
A. R. Lampis, D. J. Morrissey, M. Samuel, T. K. Murakami, and J. S. Winfield,
Phys. Rev. C \textbf{38}, 6 (1988).

\bibitem{Schu97}
Y. Schutz et al. (TAPS Collaboration), Nucl. Phys. A \textbf{622}, 404 (1997).

\bibitem{ABR}A.A. Abrikosov, L.P. Gorkov and I.E. Dzyaloshinski, Methods of Quantum Field Theory in Statistical Physics, 1963 Prentice-Hall, Inc., Englewood Cliffs, N.J., USA, 
Library of Congress Catalog Card Number 63-18808. 

\bibitem{Yin}
P.~Yin, J.~Y.~Li, P.~Wang and W.~Zuo,
%``Three-body force effect on nucleon momentum distributions in asymmetric nuclear matter within the framework of the extended Brueckner-Hartree-Fock approach,''
Phys. Rev. C \textbf{87}, 014314 (2013).


\end{thebibliography}
\end{document}